\documentclass[11pt]{article}
\usepackage[latin1]{inputenc}
\usepackage{amsmath}
\usepackage{amsfonts}
\usepackage{amssymb}
\usepackage{latexsym}
\usepackage{graphicx}
\usepackage[a4paper]{geometry}

\newtheorem{theorem}{Theorem}
\newtheorem{lemma}{Lemma}
\newtheorem{claim}{Claim}

\newenvironment{proof}{\noindent \emph{Proof.}\ }{\hfill
    $\Box$\vspace{1em}}

\newenvironment{proofcl}{\noindent \emph{Proof.}\ }{Thus the claim
holds.  \hfill $\diamond$\vspace{1em}}

\title{Characterizing path graphs by forbidden induced subgraphs}
\author{Benjamin L\'ev\^eque\thanks{ROSE, EPFL, Lausanne, Switzerland}
\and Fr\'ed\'eric Maffray\thanks{C.N.R.S., Laboratoire G-SCOP,
Grenoble, France} \and Myriam Preissmann$^\dagger$}
\begin{document}
\maketitle

\begin{abstract}
A path graph is the intersection graph of subpaths of a tree.  In
1970, Renz asked for a characterization of path graphs by forbidden
induced subgraphs.  We answer this question by determining the
complete list of graphs that are not path graphs and are minimal with
this property.
\end{abstract}

\section{Introduction}

All graphs considered here are finite and have no parallel edges and
no loop.  A \emph{hole} is a chordless cycle of length at least four.
A graph is \emph{chordal} (or \emph{triangulated}) if it contains no
hole as an induced subgraph.  Gavril~\cite{Gav74} proved that a graph
is chordal if and only if it is the intersection graph of a family of
subtrees of a tree.  In this paper, whenever we talk about the
intersection of subgraphs of a graph we mean that the \emph{vertex
sets} of the subgraphs intersect.

An \emph{interval graph} is the intersection graph of a family of
intervals on the real line; equivalently, it is the intersection graph
of a family of subpaths of a path.  An \emph{asteroidal triple} in a
graph $G$ is a set of three non adjacent vertices such that for any
two of them, there exists a path between them in $G$ that does not
intersect the neighborhood of the third.  Lekkerkerker and
Boland~\cite{LB62} proved that a graph is an interval graph if and
only if it is chordal and contains no asteroidal triple.  They derived
from this result the list of minimal forbidden subgraphs for interval
graphs.

An intermediate class is the class of path graphs.  A graph is a
\emph{path graph} if it is the intersection graph of a family of
subpaths of a tree.  Clearly, the class of path graphs is included in
the class of chordal graphs and contains the class of interval graphs.
Several characterizations of path graphs have been
given~\cite{Gav78,MW86, Ren70} but no characterization by forbidden
subgraphs was known, whereas such results exist for intersection
graphs of subpaths of a path (interval graphs~\cite{LB62}), subtrees
of a tree (chordal graphs~\cite{Gav74}), and also for directed
subpaths of a directed tree~\cite{Pan99}.

In 1970, Renz~\cite{Ren70} asked for a complete list of graphs that
are chordal and not path graphs and are minimal with this property,
and he gave two examples of such graphs.  Reference~\cite{TonGutSzw05}
extends the list of minimal forbidden subgraphs for path graphs; but
that list is incomplete.  Here we answer Renz's question and obtain a
characterization of path graphs by forbidden induced subgraphs.  We
will prove that the graphs presented in
Figures~\ref{fig:nosimpl}--\ref{fig:4cospe} are all the minimal
non-path graphs.  In other words:
\begin{theorem}
    \label{th:main}
    A graph is a path graph if and only if it does not contain any of
    $F_0, \ldots, F_{16}$ as an induced subgraph.
\end{theorem}

\section{Special simplicial vertices in chordal graphs}

In a graph $G$, a \emph{clique} is set of pairwise adjacent vertices.
Let $\mathcal Q(G)$ be the set of all (inclusionwise) maximal cliques
of $G$.  When there is no ambiguity we will write $\mathcal Q$ instead
of $\mathcal Q(G)$.

Given two vertices $u,v$ in a graph $G$, a \emph{$\{u,v\}$-separator}
is a set $S$ of vertices of $G$ such that $u$ and $v$ lie in two
different components of $G\setminus S$ and $S$ is minimal with this
property.  A set is a \emph{separator} if it is a $\{u,v\}$-separator
for some $u,v$ in $G$.  Let $\mathcal S(G)$ be the set of separators
of $G$.  When there is no ambiguity we will write $\mathcal S$ instead
of $\mathcal S(G)$.

The neighborhood of a vertex $v$ is the set $N(v)$ of vertices
adjacent to $v$.  Let us say that a vertex $u$ is \emph{complete} to a
set $X$ of vertices if $X\subseteq N(u)$.  A vertex is
\emph{simplicial} if its neighborhood is a clique.  It is easy to see
that a vertex is simplicial if and only if it does not belong to any
separator.  Given a simplicial vertex $v$, let $Q_v=N(v)\cup\{v\}$ and
$S_v=Q_v\cap N(V\setminus Q_v)$.  Since $v$ is simplicial, we have
$Q_v\in \mathcal Q$.  Remark that $S_v$ is not necessarily in
$\mathcal S$; for example, in the graph $H$ with vertices $a, b, c, d,
e$ and edges $ab, bc, cd, de, bd$, we have $S_c=\{b, d\}$ and
$\mathcal S(H)=\{\{b\}, \{d\}\}$.

A classical result~\cite{HajSur58,Ber60} (see also~\cite{Gol04})
states that, in a chordal graph $G$, every separator is a clique;
moreover, if $S$ is a separator, then there are at least two
components of $G\setminus S$ that contain a vertex that is complete to
$S$, and so $S$ is the intersection of two maximal cliques.

A \emph{clique tree} $T$ of a graph $G$ is a tree whose vertices are
the members of $\mathcal Q$ and such that, for each vertex $v$ of $G$,
those members of $\mathcal Q$ that contain $v$ induce a subtree of
$T$, which we will denote by $T^v$.  A classical result~\cite{Gav74}
states that a graph is chordal if and only if it has a clique tree.

For a clique tree $T$, the \emph{label} of an edge $QQ'$ of $T$ is
defined as $S_{QQ'}=Q\cap Q'$.  Note that every edge $QQ'$ satisfies
$S_{QQ'} \in\mathcal S$; indeed, there exist vertices $v\in Q\setminus
Q'$ and $v'\in Q'\setminus Q$, and the set $S_{QQ'}$ is a $\{v,
v'\}$-separator.  The number of times an element $S$ of $\mathcal S$
appears as a label of an edge is equal to $c-1$, where $c$ is the
number of components of $G\setminus S$ that contain a vertex complete
to $S$ \cite{Gav74,MacMac}.  Note that this number is at least one and
that it depends only on $S$ and not on $T$, so for a given $S\in
\mathcal S$ it is the same in every clique tree.

Given $X \subseteq \mathcal Q$, let $G(X)$ denote the subgraph of $G$
induced by all the vertices that appear in members of $X$.  If $T$ is
a clique tree of $G$, then $T[X]$ denotes the subtree of $T$ of
minimum size whose vertices contains $X$.  Note that if $|X|=2$, then
$T[X]$ is a path.

Given a subtree $T'$ of a clique-tree $T$ of $G$, let $\mathcal Q(T')$
be the set of vertices of $T'$ and $\mathcal S(T')$ be the set of
separators of $G(\mathcal Q(T'))$.

Dirac~\cite{Dir61} proved that a chordal graph that is not a clique
contains two non adjacent simplicial vertices.  We need to generalize
this theorem to the following.  Let us say that a simplicial vertex
$v$ is \emph{special} if $S_v$ is a member of $\mathcal S$ and is
(inclusionwise) maximal in $\mathcal S$.

\begin{theorem}
\label{special}
In a chordal graph that is not a clique, there exist two non adjacent
special simplicial vertices.
\end{theorem}
\begin{proof}
We prove the theorem by induction on $|\mathcal Q|$.  By the
hypothesis, $G$ is not a clique, so $|\mathcal Q|\geq 2$ and $\mathcal
S \neq\emptyset$.

\emph{Case 1: $\mathcal S$ has only one maximal element $S$.} Let
$Q,Q'$ be two maximal cliques such that $Q\cap Q'=S$.  Let $v\in
Q\setminus Q'$ and $v'\in Q'\setminus Q$.  The set $S$ is the only
maximal separator and it does not contain $v$ or $v'$.  So $v$ and
$v'$ do not belong to any element of $\mathcal S$, and so they are
simplicial and $S_v=S_{v'}=S$, so they are special.

\emph{Case 2: $\mathcal S$ has two distinct maximal elements $S,S'$}.
So $|\mathcal Q|\ge 3$.  Let $T$ be a clique tree of $G$.  Let $Q_1,
Q_2, Q_1', Q_2'$ be members of $\mathcal Q$ such that $S=S_{Q_1Q_2}$,
$S'=S_{Q_1'Q_2'}$, and $Q_2,Q_1,Q'_1,Q'_2$ appear in this order along
the path $T[Q_2, Q_1, Q'_1, Q'_2]$ (possibly $Q_1=Q'_1$).  Let $Y$ be
the subtree of $T\setminus Q_1$ that contains $Q_2$, and let $Z$ be
the tree that consists of $Y$ plus the vertex $Q_1$ and the edge
$Q_1Q_2$.  The subtree $Z$ does not contain $Q_2'$, so $G (\mathcal
Q(Z))$ has strictly fewer maximal cliques than $G$; and $G$ is not a
clique.  By the induction hypothesis, there exist two non adjacent
simplicial vertices $v, w$ of $G(\mathcal Q(Z))$ such that $S_{v},
S_{w}$ are maximal elements of $\mathcal S(Z)$.  At most one of $v,w$
is in $Q_1$ since they are not adjacent, say $v$ is not in $Q_1$.  We
claim that $v$ is a simplicial vertex of $G$ and that $S_v$ is a
maximal element of $\mathcal S$.  Vertex $v$ does not belong to any
element of $\mathcal S(Z)$.  If it belongs to an element of $\mathcal
S\setminus \mathcal S(Z)$, then it must also belong to $Q_1\cap Q_2 =S
\in\mathcal S(Z)$, a contradiction.  So $v$ does not belong to any
element of $\mathcal S$ and so it is a simplicial vertex of $G$.  The
set $S_v$ is a maximal element of $\mathcal S(Z)$.  If it is not a
maximal element of $\mathcal S$, then it is included in $S \in\mathcal
S(Z)$, a contradiction.  So $v$ is a special simplicial vertex of $G$.
Likewise, let $Y'$ be the subtree of $T\setminus Q_1'$ that contains
$Q_2'$, and let $Z'$ be the tree that consists of $Y'$ plus the vertex
$Q_1'$ and the edge $Q_1'Q_2'$.  Just like with $v$, we can find a
simplicial vertex $v'$ of $G(\mathcal Q(Z'))$ not in $Q_1'$ that is a
simplicial vertex of $G$ with $S_{v'}$ being a maximal element of
$\mathcal S$.  Vertices $v$ and $v'$ are not adjacent since $S$ is a
$\{v, v'\}$-separator.  So $v$ and $v'$ are the desired vertices.
\end{proof}

Algorithms LexBFS~\cite{RTL76} and MCS~\cite{TY84} are linear time
algorithms that were developed to find a simplicial vertex in a
chordal graph.  But a simplicial vertex found by these algorithms is
not necessarily special.  For example, on the graph with vertices
$a,b,c,d,e,f$ and edges $ab,bc, cd, eb, ec, fb, fc$, every application
of LexBFS or MCS will end on one of simplicial vertices $a, d$, which
are not special.  The proof of Theorem~\ref{special} can be turned
into a polynomial time algorithm to find a special simplicial vertex
in a chordal graph.  We do not know how to find such a vertex in
linear time.

\section{Forbidden induced subgraphs}

A \emph{clique path tree} $T$ of $G$ is a clique tree of $G$ such
that, for each vertex $v$ of $G$, the subtree $T^v$ induced by cliques
that contain $v$ is a path.  Gavril \cite{Gav78} proved a graph is a
path graph if and only if it has a clique path tree.

\begin{figure}[e]
      \centering
\begin{tabular}{c}
\includegraphics[scale=0.5]{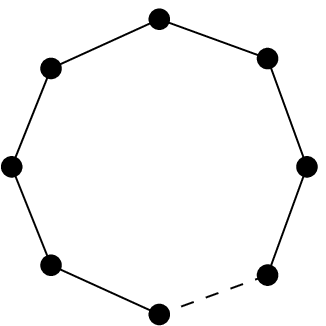} \\
$F_0(n)_{n\geq 4}$ \\
\end{tabular}
\caption{Forbidden subgraphs with no simplicial vertices}
\label{fig:nosimpl}
    \end{figure}

\begin{figure}[e]
      \centering
\begin{tabular}{ccccc}
\includegraphics[scale=0.5]{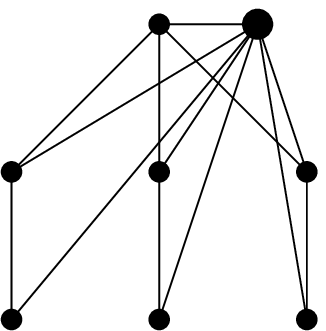} \ &\
\includegraphics[scale=0.5]{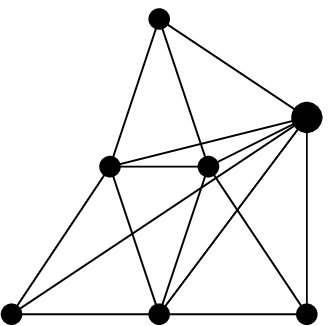} \ &\
\includegraphics[scale=0.5]{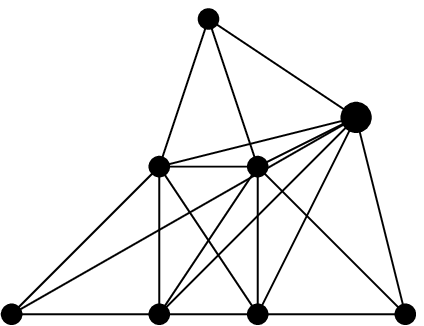} \ &\
\includegraphics[scale=0.5]{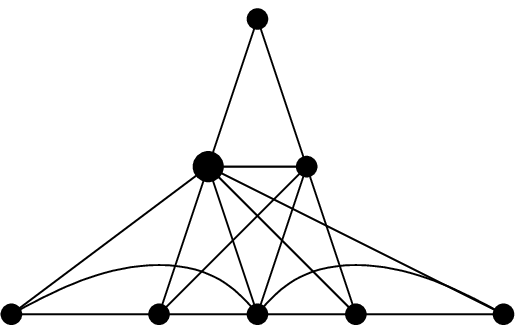} \ &\
\includegraphics[scale=0.5]{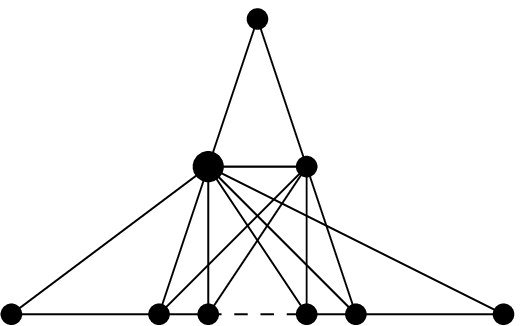} \\
$F_1$ & $F_2$ & $F_3$ & $F_4$ & $F_5(n)_{n\geq 7}$
\end{tabular}
      \caption{Forbidden subgraphs with a universal vertex}
\label{fig:univ}
    \end{figure}

\begin{figure}[e]
      \centering
\begin{tabular}{ccccc}
\includegraphics[scale=0.5]{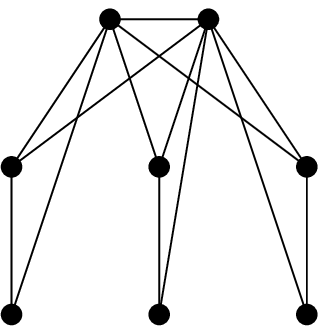} \ &\
\includegraphics[scale=0.5]{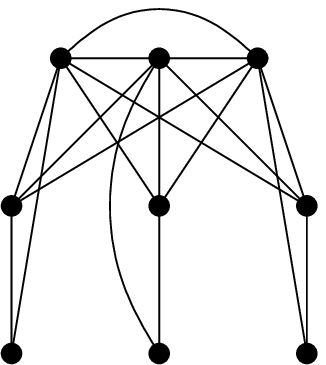} \ &\
\includegraphics[scale=0.5]{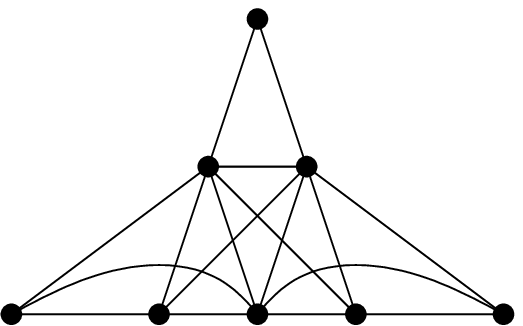} \ &\
\includegraphics[scale=0.5]{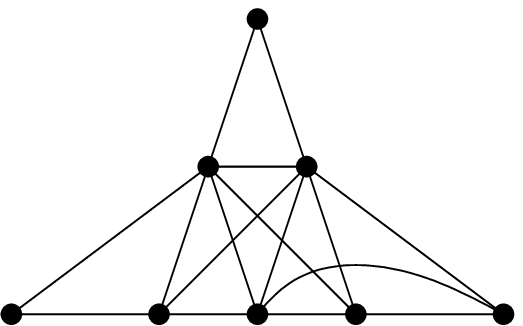}  \ &\
\includegraphics[scale=0.5]{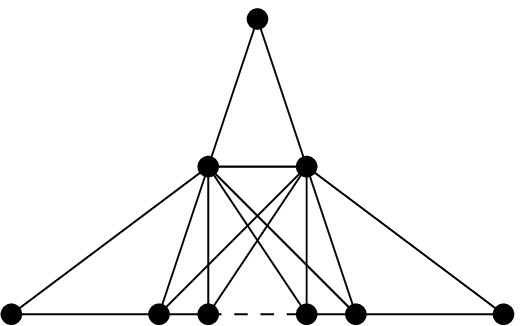} \\
$F_6$ & $F_7$ & $F_8$ & $F_9$ & $F_{10}(n)_{n\geq 8}$ \\
\end{tabular}
      \caption{Forbidden subgraphs with no universal vertex and exactly
      three simplicial vertices}
\label{fig:3simpl}
    \end{figure}

\begin{figure}[e]
      \centering
\begin{tabular}{ccccc}
\includegraphics[scale=0.5]{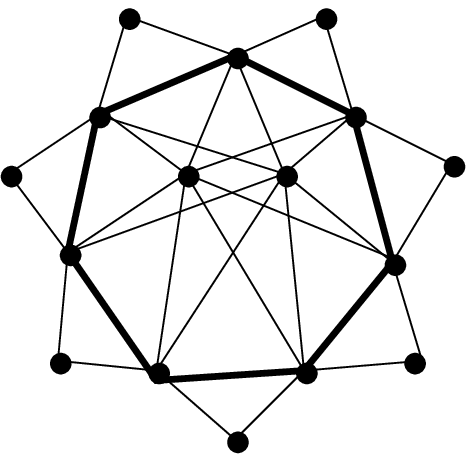}\ &\
\includegraphics[scale=0.5]{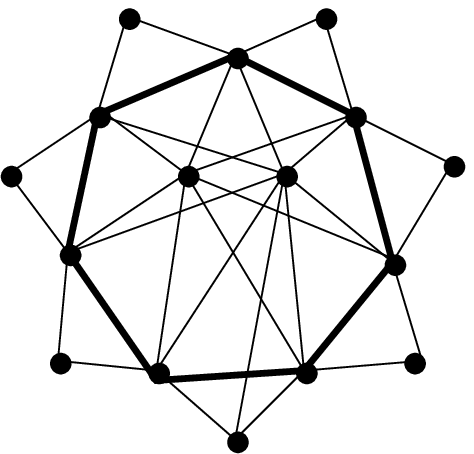} \ &\
\includegraphics[scale=0.5]{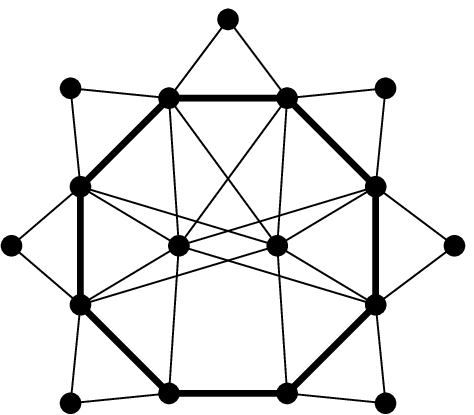}\ &\
\includegraphics[scale=0.5]{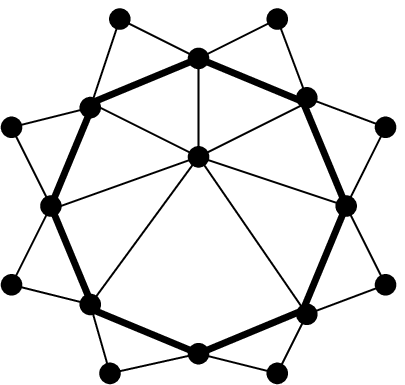}  \ &\
\includegraphics[scale=0.5]{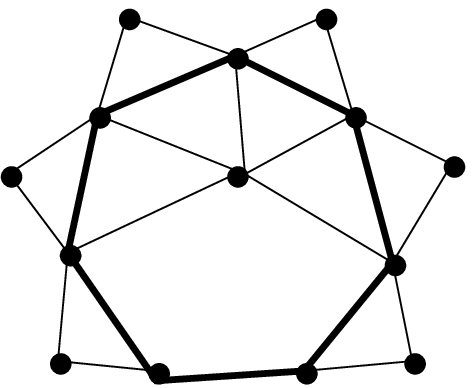} \\
$F_{11}(4k)_{k\geq 2}$ & $F_{12}(4k)_{k\geq 2}$ &
$F_{13}(4k+1)_{k\geq 2}$ & $F_{14}(4k+1)_{k\geq 
2}$ & $F_{15}(4k+2)_{k\geq 2}$ \\
\end{tabular}
\caption{Forbidden subgraphs with at least one simplicial vertex that
    is not co-special. (bold edges form a clique)}
\label{fig:4simpl}
    \end{figure}

\begin{figure}[e]
      \centering
\begin{tabular}{c}
\includegraphics[scale=0.5]{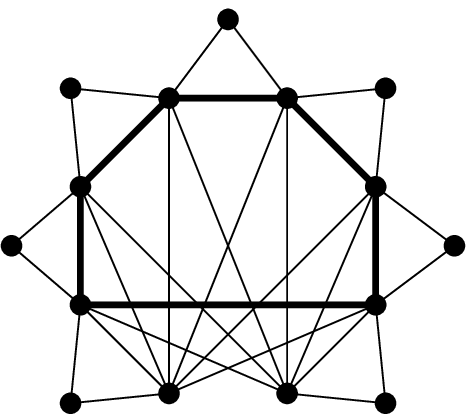} \\
$F_{16}(4k+3)_{k\geq 2}$ \\
\end{tabular}
\caption{Forbidden subgraphs with $\geq 4$ simplicial vertices that
    are all co-special. (bold edges form a clique)}
\label{fig:4cospe}
    \end{figure}

Consider graphs $F_0, \ldots, F_{16}$ presented in
Figures~\ref{fig:nosimpl}--\ref{fig:4cospe}.  Let us make a few
remarks about them.  Each graph in Figure~\ref{fig:univ} is obtained
by adding a universal vertex to some minimal forbidden subgraph for
interval graphs.  Clearly, in a path graph the neighborhood of every
vertex is an interval graph; so $F_1, \ldots, F_5$ are not path
graphs.  Graphs $F_{10}(n)_{n\geq 8}$ are also forbidden in interval
graphs.  Graphs $F_6$ and $F_{10}(8)$ are from Renz~\cite[Figures~1
and~5]{Ren70}.  For $i\in \{0, 1, 3, 4, 5, 6, 7, 9, 10, 13, 15, 16\}$,
Panda~\cite{Pan99} proved that $F_i$ is a minimal non directed path
graph, so $F_i\setminus x$ is a directed path graph for every vertex
$x$ (obviously every directed path graph is a path graph).  In general
we have the following:
\begin{theorem}
    \label{th:minimal}
    $F_0,\ldots,F_{16}$ are minimal non path graphs.
\end{theorem}
\begin{proof}
Clearly, $F_0$ is a minimal non path graph.  For the other graphs, we
prove the theorem in one case and then show how the same arguments can
be applied to all cases.

Consider $F=F_{11}(4k)$, $k\ge 2$; see Figure~\ref{fig:4simpl}.  Name
its vertices such that $u_1, \ldots, u_{2k-1}$ are the simplicial
vertices of degree $2$, clockwise; $v_{j-1}, v_j$ are the two
neighbors of $u_j$ ($j=1, \ldots, {2k-1}$), with subscripts modulo
$2k-1$; and $a,b$ are the remaining vertices.  Let $Q_j$ be the
maximal clique that contains $u_j$ ($j=1, \ldots, {2k-1}$), and call
these $2k-1$ cliques ``peripheral''.  Let $R_a=\{a, v_1, \ldots,
v_{2k-1}\}$ and $R_b=\{b, v_1, \ldots, v_{2k-1}\}$ be the maximal
cliques that contain respectively $a$ and $b$, and call these two
cliques ``central''.  Thus $\mathcal Q(F)=\{R_a, R_b, Q_1, \ldots,
Q_{2k-1}\}$.  Since $F$ is chordal, it admits a clique tree.  Let $T$
be any clique tree of $F$.  Then $R_a$ and $R_b$ are adjacent in $T$
(for otherwise, there would be at least one interior vertex $Q$ on the
path $T[R_a, R_b]$, so we should have $R_a\cap R_b\subseteq Q$, but no
member $Q$ of $\mathcal Q(F)\setminus\{R_a, R_b\}$ satisfies this
inclusion).  By the same argument, each $Q_j$ ($j=1, \ldots, 2k-1$)
must be adjacent to $R_a$ or $R_b$ in $T$.  Suppose that we are trying
to build a clique path tree $T$ for $F$.  By symmetry, we may assume
that $Q_1$ is adjacent to $R_a$.  Then, for $j=2, \ldots, 2k-2$
successively, $Q_j$ must be adjacent to $R_b$ (if $j$ is even) and to
$R_a$ (if $j$ is odd) in $T$, for otherwise, for some $v\in\{v_{j-1},
v_j\}$ the subtree $T^v$ induced by the cliques that contain $v$ would
not be a path.  Note that in this fashion we obtain a clique path tree
$T'$ of $F\setminus u_{2k-1}$.  Now if $Q_{2k-1}$ is adjacent to
$R_a$, then the subtree $T^{v_{2k-1}}$ is not a path, and if if
$Q_{2k-1}$ is adjacent to $R_b$, then the same holds for
$T^{v_{2k-2}}$.  This shows that $F$ is not a path graph.

Now consider any vertex $x$ of $F$.  If $x$ is one of the $u_j$'s,
then by symmetry we may assume that $x=u_{2k-1}$, and we have seen
above that $F\setminus x$ is a path graph with clique path tree $T'$.
Suppose that $x$ is one of the $v_j$'s, say $x=v_{2k-1}$.  Then by
adding vertex $Q_{2k-1}$ and edge $Q_{2k-1}R_a$ to $T'$, it is easy to
see that we obtain a clique path tree of $F\setminus x$.  Finally,
suppose that $x$ is one of $a,b$, say $x=b$.  Then the tree with
vertices $R_a, Q_1, \ldots, Q_{2k-1}$ and edges $R_aQ_1, \ldots,
R_aQ_{2k-1}$ is a clique path tree of $F\setminus x$.  So $F$ is a
minimal non path graph.

When $F$ is any other $F_i$ ($i=1, \ldots, 16$), the same arguments
apply as follows.  For $i=1, \ldots, 10$, call peripheral the three
cliques that contain a simplicial vertex.  For $i=11, \ldots, 16$,
call peripheral the cliques that contain a simplicial vertex of degree
$2$, plus, in the case of $F_{12}$, the clique that contain the bottom
simplicial vertex (which has degree $3$).  Call central all other
maximal cliques.  Then it is easy to prove, as above, that the central
cliques must form a subpath in any clique tree of $F$, and all the
peripheral cliques except one can be appended to either end of that
subpath, but whichever way this is done, when the last clique is
appended, the subtree $T^v$ is not a path for some vertex $v$ of $F$.
Moreover, when any vertex $x$ is removed, it is possible to build a
clique path tree for $F\setminus x$.
\end{proof}

\section{Co-special simplicial vertices}

Let us say that a simplicial vertex $v$ is \emph{co-special} if $S_v$
is a separator such that $G\setminus S_v$ has exactly two components.
Note that in that case $S_v$ is a minimal element of $\mathcal S$ and
it appears exactly once as a label of any path tree of $G$.

\begin{lemma}
\label{cospecial}
Let $G$ be a minimal non path graph.  Then either $G$ is one of
$F_{11},\ldots F_{15}$ or every simplicial vertex of $G$ is
co-special.
\end{lemma}
\begin{proof}
Suppose on the contrary that $G$ is a minimal non path graph,
different from $F_{11},\ldots F_{15}$, and there is a simplicial
vertex $q$ of $G$ that is not co-special.  All simplicial vertices of
$F_0,\ldots F_{10}, F_{16}$ are co-special, so $G$ is not any of these
graphs; moreover it does not contain any of them strictly (for
otherwise $G$ would not be minimal).  Therefore $G$ contains none of
$F_0, \ldots, F_{16}$.

Let $T_0$ be a clique path tree of $G\setminus q$.  Let $Q'\in
\mathcal Q(G\setminus q)$ be such that $S_q\subseteq Q'$.  If
$Q'=S_q$, then we can add $q$ to $Q'$ to obtain a clique path tree of
$G$, a contradiction.  So $Q'\neq S_q$, and $S_q \in \mathcal S$ (as
there is a vertex $q'\in Q'\setminus S_q$ and $S_q$ is a $\{q,
q'\}$-separator).

Let $T'$ be the maximal subtree of $T_0$ that contains $Q'$ and such
that no label of the edges of $T_0$ is included in $S_q$.  Remark that
$T'$ plus vertex $Q$ and edge $QQ'$ is a clique tree of $G(\mathcal
Q(T')\cup \{Q\})$ (but not necessarily a clique path tree), and in
that tree only one label is included in $S_q$.  Since $q$ is not
co-special, there is an edge of $T_0$ whose label is included in
$S_q$, and so $T'$ has strictly fewer vertices than $T_0$.  So
$G(\mathcal Q(T')\cup \{Q\})$ is a path graph.  Let $T$ be a clique
path tree of this graph.

We claim that $Q$ is a leaf of $T$.  If not, then there are at least
two labels of $T$ that are included in $S_q$, which contradicts the
definition of $T'$ (the number of times a label appears in a clique
tree is constant).

Let $T_1,\ldots, T_\ell$ be the subtrees of $T_0\setminus T'$
($\ell\geq 1$).  For $1\leq i\leq \ell$, let $Q_iQ_i'$ be the edge
between $T_i$ and $T'$ with $Q_i\in T_i$ and $Q_i'\in T'$.  Note that
$Q_1, \ldots, Q_\ell$ are pairwise disjoint (but $Q', Q_1', \ldots,
Q_\ell'$ are not necessarily pairwise disjoint).  Let $S_i=Q_i\cap
Q_i'$ and $v_i\in Q_i\setminus Q_i'$.  Let ${\mathcal H}= (V_{\mathcal
H}, E_{\mathcal H})$ be the intersection graph of $S_1, \ldots,
S_\ell$, that is, $V_{\mathcal H}= \{ S_1, \ldots, S_\ell\}$ and
$E_{\mathcal H}=\{ S_i S_j\ |\ S_i\cap S_j\neq \emptyset\}$.

\begin{claim}
$\mathcal H$ contains no odd cycle.
\end{claim}
\begin{proofcl}
Suppose on the contrary, without loss of generality, that
$S_1$-$\cdots$-$S_p$-$S_1$ is an odd cycle in $\mathcal H$, with
length $p=2r+1$ ($r\geq 1$).  Let $I_j= S_{j}\cap S_{j+1}$ ($j=1,
\ldots, p$), with $S_{p+1}=S_1$.  Suppose that for some $j\neq k$ we
have $I_j\cap I_k \neq \emptyset$; then there is a common vertex in
the cliques $Q_{j}, Q_{j+1}, Q_{k}, Q_{k+1}$, and the number of
different cliques among these is at least three, which contradicts the
fact that $T_0$ is a clique path tree as these three cliques do not
lie on a common path of $T_0$.  For $1\leq j\leq p$, let $s_j\in I_j$.
By the preceding remark, the $s_j$'s are pairwise distinct.  By the
definition of $T'$, we have $S_j\subseteq S_q$ for each $1\leq j\leq
p$, so the $s_j$'s are all in $Q$ and $Q'$.  Let $q'\in Q' \setminus
Q$.  Let us consider the subgraph induced by $q, q', v_1, \ldots, v_p,
s_1, \ldots, s_p$.  Each of the non-adjacent vertices $q$ and $q'$ is
adjacent to all of the clique formed by the $s_j$'s.  Each vertex
$v_j$ is adjacent to $s_{j-1}$ and $s_j$ (with $s_0=s_p$) and not to
any other $s_i$ or to $q$.  Vertex $q'$ can have at most two neighbors
among the $v_j$'s.  If $q'$ has zero or one neighbor among them, then
$q, q', v_1, \ldots, v_p, s_1, \ldots, s_p$ induce respectively
$F_{11}(4r+4)_{r\geq 1}$ or $F_{12}(4r+4)_{r\geq 1}$.  If $q'$ has two
consecutive neighbors $v_{j}, v_{j+1}$, then $q, q', v_{j}, v_{j+1},
s_{j-1}, s_{j}, s_{j+1}$ induce $F_2$.  If $q'$ has two
non-consecutive neighbors $v_{j}, v_{k}$, then we can assume that
$1\leq j<j+1<k\leq p$ and $k-j$ is odd, $k-j=2s+1$ with $s\geq 1$, and
then $q, q', v_{j}, \ldots, v_{k}, s_{j}, \ldots, s_{k-1}$ induce
$F_{14}(4s+5)_{s\geq1}$.  In all cases we obtain a contradiction.
\end{proofcl}

By the preceding claim, $\mathcal H$ is a bipartite graph.

For $1\leq i\leq \ell$, let $\mathcal R_i=\{S\in \mathcal S(T')\ |\
S_i\cap S \neq \emptyset\ \textrm{and}\ S_i\setminus S\neq
\emptyset\}$.  Let $X=\{S_i\ |\ \mathcal R_i\neq \emptyset\}$.
%
\begin{claim}
$\mathcal H$ contains no odd path between two vertices in $X$.
\end{claim}
\begin{proofcl}
Suppose on the contrary, without loss of generality, that
$S_1$-$\cdots$-$S_p$ is an odd path in $\mathcal H$ between two
vertices $S_1, S_p$ of $X$ (with $p=2k$, $k\geq 1$), and assume that
$p$ is minimum with this property.  By the minimality, all interior
vertices $S_{j}$ ($1<j<p$) are not in $X$.  For $1\leq j<p$, let $s_j$
be a vertex in $S_{j}\cap S_{j+1}$.  As in the preceding claim, the
$s_j$'s are pairwise distinct and lie in $Q$ and $Q'$.  Let $P$ be the
path $T'[Q_1', Q_2']$.  If $p\neq 2$, then $S_2$ is not in $X$, so
$Q_3'=Q_1'$, for otherwise $T_0^{s_2}$ would not be a path; then $S_3$
is not in $X$, so $Q_{4}'=Q_2'$, and so on.  Thus the two extremities
of $P$ are $Q_1'=Q_3'= \cdots = Q_{p-1}'$ and $Q_2'=Q_{4}'= \cdots =
Q_p'$.  Since $S_1$ and $S_p$ are in $X$, the sets $\mathcal R_1,
\mathcal R_p$ are non empty.


Let $L_1$ be the closest vertex to $Q_1'$ in $P$ such that there
exists an edge incident to $L_1$ with label in $\mathcal R_1$, and let
$L_1K_1$ be such an edge and $R_1$ be its label (such an edge exists
because $\mathcal R_1\neq\emptyset$).  Similarly, let $L_p$ be the
closest vertex to $Q_p'$ in $P$ such that there exists an edge
incident to $L_p$ with label in $\mathcal R_p$, and let $L_pK_p$ be
such an edge and $R_p$ be its label.  So $S_1 \subseteq L_1$,
$S_1\nsubseteq K_1$ and $S_p \subseteq L_p$, $S_p \nsubseteq K_p$.
Each of $K_1, K_p$ may be in $P$ or not.  Since $T'$ is a clique path
tree, $Q'$ lies between $Q_1'$ and $L_1$ and between $L_p$ and $Q_p'$
along $P$.  So $Q_1', L_p, Q', L_1, Q_p'$ lie in this order on $P$,
and $S_1$ is included in all labels between $Q_1'$ and $L_1$ in $P$,
and $S_p$ is included in all labels between $Q_p'$ and $L_p$ in $P$.

Let $v_0 \in K_1\setminus L_1$ and $v_{p+1} \in K_p\setminus L_p$.
Since $T_0$ is a clique path tree, $v_0$ and $v_{p+1}$ are
distinct from $v_1, \ldots, v_p$ and not adjacent to $q$.

Let $s_0 \in S_1\cap R_1$ and $s_p \in S_p\cap R_p$.  Then $v_0$ and
$s_0$ are adjacent, and $v_{p+1}$ and $s_p$ are adjacent.  Since $T_0$
is a clique path tree, if $K_1$ or $K_p$ is not in $P$, then $s_0$ and
$s_p$ are different from each other, from $s_1, \ldots, s_{p-1}$ and
from $v_0, \ldots, v_{p+1}$.  Furthermore, if $K_1$ is not in $P$,
then $v_0$ is not adjacent to any of $s_1, \ldots, s_p$; and if $K_p$
is not in $P$, then $v_{p+1}$ is not adjacent to any of $s_0, \ldots,
s_{p-1}$.

Let $s_0' \in S_1\setminus R_1$ and $s_p' \in S_p\setminus R_p$.  Then
$v_0$ and $s_0'$ are not adjacent, and $v_{p+1}$ and $s_p'$ are not
adjacent.  Since $T_0$ is a clique path tree, if $K_1$ or $K_p$ is in
$P$, then $s_0'$ and $s_p'$ are different from each other, from $s_1,
\ldots, s_{p-1}$ and from $v_0, \ldots, v_{p+1}$.  Furthermore, if
$K_1$ is in $P$, then $v_0$ is adjacent to $s_p'$ and to $s_1, \ldots,
s_p$; and if $K_p$ is in $P$, then $v_{p+1}$ is adjacent to $s_0'$ and
to $s_0, \ldots, s_{p-1}$.

Note that the set $\{q, s_0', s_0, s_1, s_2, \ldots, s_p, s_p'\}$
induces a clique in $G$.  Moreover, $v_1$ is adjacent to $s'_0$, $v_p$
is adjacent to $s'_p$, for $i= 1, \ldots, p$, $v_{i}$ is adjacent to
$s_{i-1}$ and $s_{i}$, and there is no other edge between $v_1,
\ldots, v_p$ and that clique.

Suppose that $K_1=K_p$.  Then $L_1= L_p=Q'$ and $K_1$ is not in $P$.
By the definition of $T'$, there exists $y\in R_1\setminus S_q$.
Vertex $y$ is distinct from all $s_i$'s as it is not in $S_q$, and it
is adjacent to all of $v_0, s_0, \ldots, s_p$ and to none of $q, v_1,
\ldots, v_p$.  Then $q, y, v_0, \ldots, v_p, s_0, \ldots , s_p$ induce
$F_{12}(4k+4)_{k\geq1}$, a contradiction.  So $K_1\neq K_p$, and $v_0$
and $v_{p+1}$ are distinct non adjacent vertices.  We can choose
vertices $x_1, \ldots, x_r$ ($r\geq 1$) not in $S_q$ and on the labels
of $T'[K_1, K_p]$ such that $v_0$-$x_1$-$\ldots$-$x_r$-$v_{p+1}$ is a
chordless path in $G$.  Vertices $x_1, \ldots, x_r$ are distinct from
and adjacent to $s_0', s_p', s_0, \ldots, s_p$, and they are distinct
from and not adjacent to any of $v_1, \ldots v_p$.

Suppose that $L_1 = Q_p'$ and $L_p = Q_1'$.  Then $K_1$ and $K_p$ are
not in $P$.  If $r=1$, then $q, v_0, \ldots, v_{p+1},$ $s_0, \ldots,
s_p, x_1$ induce $F_{14}(4k+5)_{k\geq1}$.  If $r=2$, then $q, v_0,
\ldots, v_{p+1},$ $s_0, \ldots, s_p, x_1, x_2$ induce
$F_{15}(4k+6)_{k\geq1}$.  If $r\geq 3$, then $q, v_0, v_{p+1}, s_0,
s_p, x_1, \ldots, x_r$ induce $F_{10}(r+5)_{r\geq3}$, a contradiction.

Suppose now that $L_1 \neq Q_p'$ and $L_p = Q_1'$.  Then $K_p$ is not
in $P$ and we may assume that $K_1$ is in $P$.  If $r=1$, then $q,
v_0, \ldots, v_{p+1},$ $s_0', s_1 \ldots, s_p, x_1$ induce
$F_{13}(4k+5)_{k\geq1}$.  If $r\geq 2$, then $q, v_0, v_{p+1}, x_1,
\ldots, x_r, s_0', s_p$ induce $F_{5}(r+5)_{r\geq2}$, a contradiction.

Suppose finally that $L_1 \neq Q_p'$ and $L_p \neq Q_1'$.  Then we may
assume that $K_1$ and $K_p$ are in $P$.  If $r=1$, then $q, v_0,
v_{p+1}, s_0', s_1, s_p', x_1$ induce $F_2$.  If $r=2$, then $q, v_0,
v_{p+1}, s_0', s_1, s_p', x_1, x_2$ induce $F_3$.  If $r\geq 3$, then
$q, v_0, v_{p+1}, x_1, \ldots, x_r, s_0', s_p'$ induce
$F_{10}(r+5)_{r\geq3}$, a contradiction.
\end{proofcl}

By the preceding two claims, $\mathcal H$ is a bipartite graph $(A, B,
E_{\mathcal H})$ such that $X\subseteq A$.  Now all the subtrees $T_i$
can be linked to $T$ to get a clique path tree of $G$ as follows.  For
each $S_i\in A$, we add an edge $QQ_i$ between $T$ and $T_i$.  This
creates a clique path tree on the corresponding subset of cliques
because $A$ is a stable set of $\mathcal H$ and $Q$ is a leaf of $T$.
For each $S_i\in B$, let $Q_i''\in \mathcal Q(T)$ be such that
$Q_i''\cap S_i \neq \emptyset$ and the length of $T[Q, Q_i'']$ is
maximal.  Since $S_i\in B$, we have $\mathcal R_i=\emptyset$, so
$S_i\subseteq Q_i''$ and we can add an edge $Q_i''Q_i$ between $T$ and
$T_i$.  This creates a clique path tree of $G$ because $B$ is a stable
set of $\mathcal H$ and by the definition of $Q_i''$, a contradiction.
\end{proof}

\section{Characterization of path graphs}

In this section we prove the main theorem, that is, path graphs are
exactly the graphs that do not contain any of $F_0, \ldots, F_{16}$.
We could not find a characterization similar to the one found by
Lekkerkerker and Boland~\cite{LB62} for interval graphs (``an interval
graph is a chordal graph with no asteroidal triple'').  We know that
in a path graph, the neighborhood of every vertex contains no
asteroidal triple; but this condition is not sufficient.  So we prove
directly that a graph that does not contain any of the excluded
subgraphs is a path graph.

\begin{lemma}
\label{lem:PAT}
In a graph that does not contain any of $F_0, \ldots, F_5, F_{10}$,
the neighborhood of every vertex does not contain an asteroidal
triple.
\end{lemma}

\begin{proof}
    It suffices to check that when a universal vertex is added to a
    minimal forbidden induced subgraph for interval graphs
    (\cite{LB62}), then one obtains a graph that contains one of $F_0,
    \ldots, F_{5}, F_{10}$.  The easy details are left to the reader.
\end{proof}

Given three non adjacent vertices $a, b, c$, we say that $a$ is the
\emph{middle} of $b, c$ if every path between $b$ and $c$ contains a
vertex from $N(a)$.  If $a, b, c$ is not an asteroidal triple, then at
least one of them is the middle of the others.

\begin{lemma}
    \label{lem:middle}
In a chordal graph $G$ with clique tree $T$, a vertex $a$ is the
middle of two vertices $b, c$ if and only if for all cliques $Q_b$ and
$Q_c$ such that $b\in Q_b$ and $c\in Q_c$, there is an edge of the
path $T[Q_b, Q_c]$ such that $a$ is complete to its label.
\end{lemma}

\begin{proof}
Suppose that $a$ is the middle of $b, c$.  Let $Q_b$ and $Q_c$ be
cliques such that $b\in Q_b$ and $c\in Q_c$, and suppose there is no
edge of $T[Q_b, Q_c]$ such that $a$ is complete to its label.  For
each edge on $T[Q_b, Q_c]$, one can select a vertex that is not
adjacent to $a$.  Then the set of selected vertices forms a path from
$b$ to $c$ that uses no vertex from $N(a)$, a contradiction.

Suppose now that for all cliques $Q_b$ and $Q_c$ with $b\in Q_b$ and
$c\in Q_c$, there is an edge of the path $T[Q_b, Q_c]$ such that $a$
is complete to its label.  Suppose that there exists a path
$x_0$-$\cdots$-$x_r$, with $b=x_0$ and $c=x_r$ and none of the $x_i$'s
is in $N(a)$.  We may assume that this path is chordless.  For $1 \leq
i \leq r$, let $Q_i$ be a maximal clique containing $x_{i-1}, x_i$.
Then $Q_1, \ldots, Q_r$ appear in this order along a subpath of $T$.
On each $T[Q_i, Q_{i+1}]$ ($1\leq i\leq r-1$), vertex $a$ is not
adjacent to $x_i$, so $a$ is not complete to any label of $T[Q_1,
\ldots, Q_r]$, but $Q_1$ contains $b$ and $Q_r$ contains $c$, a
contradiction.
\end{proof}

Now we are ready to prove the main theorem.  Part of the proof has be
done in the previous section.  Lemma~\ref{cospecial} deals with the
case where there exists a simplicial vertex that is the middle of two
other vertices; now we have to look at the case where all simplicial
vertices are not the middle of any pair of vertices.

\paragraph*{Proof of Theorem~\ref{th:main}}
\setcounter{claim}{0}

By Theorem~\ref{th:minimal}, a path graph does not contain any of
$F_0, \ldots, F_{16}$.  Suppose now that there exists a graph $G$ that
does not contain any of $F_0, \ldots, F_{16}$ and is a minimal non
path graph.  Since $G$ contains no $F_0$, it is chordal.  By
Theorem~\ref{special}, there is a special simplicial vertex $q$ of
$G$.  By Lemma~\ref{cospecial}, $q$ is co-special.  Let $Q=Q_q$ and
$S_Q=S_q\in \mathcal S$.  Let $T_0$ be a clique path tree of
$G(\mathcal Q\setminus Q)$.  Let $Q'\in \mathcal Q\setminus Q$ be such
that $S_Q\subseteq Q'$.  We add the edge $QQ'$ to $T_0$ to obtain a
clique tree $T_0'$ of $G$.

\begin{claim}\label{clq}
     For all non-adjacent vertices $u,w\notin Q$, there exists a path
     between $u$ and $v$ that avoids the neighbourhood of $q$.
\end{claim}
\begin{proofcl}
Suppose the contrary.  Let $U, W\in \mathcal Q$ be such that $u\in U$
and $w\in W$.  We have $U\neq W$ since $u, w$ are not adjacent.  By
Lemma~\ref{lem:middle}, there is an edge of $T_0[U, W]$ whose label is
included in $S_Q$, contradicting that $q$ is co-special.
\end{proofcl}

For each clique $L \in \mathcal Q\setminus \{Q, Q'\}$, let $L'$ be the
neighbor of $L$ along $T_0[L, Q']$.  Let $S_L=L\cap L'$.  Let
$\mathcal S_L$ be the set of labels of edges incident to $L$ in $T_0$.
Let $\overline L$ be the clique in $T_0[L, Q']\setminus \{Q'\}$ such
that $S_{\overline L}\subseteq S_L$ and no other edge of
$T_0[\overline L, Q']$ has a label included in $S_L$.  (Possibly
$\overline L = L$.)

Let $\mathcal L$ be the set of cliques $L$ of $\mathcal Q\setminus
\{Q, Q'\}$ such that no element of $\mathcal S_L\setminus S_L$
contains~$S_{\overline L}$.  For each clique $L \in \mathcal L$, we
define a subtree $T_L$ of $T_0'$, where $T_L$ is the biggest subtree
of $T_0'$ that contains $Q'$ and for which no label is included in
$S_L$.  Note that $\overline L'$ is in $T_L$ and $\overline L$ is not
in $T_L$.  Since $q$ is special and co-special we have $S_Q\nsubseteq
S_L$, so $T_L$ contains $Q$.

   \begin{claim}
   \label{claim:L'}
   For each clique $L \in \mathcal L$ we have $L'\in T_L$.
   \end{claim}
\begin{proofcl}
Suppose on the contrary that $L'\notin T_L$.  Then $\overline L\neq
L$.  When we remove the edges $LL'$ and $\overline L \overline L'$
from $T_0'$, there remain three subtrees $T_1, T_2, T_3$, where $T_1$
is the subtree that contains $L$, $T_2$ is the subtree that contains
$L'$ and $\overline L$, and $T_3$ is the subtree that contains
$\overline L', Q', Q$.  Let $T_4$ be the tree formed by $T_1, T_3$
plus the edge $L\overline L'$.  Then, since $S_{\overline L}\subseteq
S_L$, $T_4$ is a clique tree of $G(\mathcal Q(T_4))$.  The set
$\mathcal Q(T_4)$ contains strictly fewer maximal cliques than
$\mathcal Q$, so there exists a clique path tree $T_5$ of $G(\mathcal
Q(T_4))$.  Label $S_{\overline L}$ is on the edge $L\overline L'$ of
$T_4$, so it is also a label of $T_5$.  Consequently there is an edge
$LL''$ of $T_5$ with a label $R$ such that $S_{\overline L} \subseteq
R \subseteq L$.  (Possibly $L''=\overline L'$).  Suppose that $R\neq
S_{\overline L}$.  Then there is an edge of $T_1$ or $T_3$ with label
$R$.  But no label of $T_1$ can be $R$ by the definition of $\mathcal
L$; and all the labels of $T_3$ that are included in $L$ are also
included in $S_{\overline L}$, so no label of $T_3$ can be $R$, a
contradiction.  So $R=S_{\overline L}$.  Now if we remove the edge
$LL''$ from $T_5$ and replace it by the subtree $T_2$ and edges $LL'$
and $\overline L L''$, we obtain a clique path tree of $G$, a
contradiction.
\end{proofcl}


Let $\mathcal L^*$ be the set of all $L\in \mathcal L$ such that $T_L$
is a strict subtree of $T_0'\setminus L$.  For every vertex $x$ of
$G(\mathcal Q\setminus Q)$ let $T_0^x$ be the subtree of $T_0$ induced
by the cliques that contain $x$.  Recall that $T_0^x$ is a path
because $T_0$ is a clique path tree.  Let $A$ be the set of vertices
$a$ of $Q$ such that $Q'$ is a vertex of $T_0^a$ that is not a leaf.
Then $A$ is not empty, for otherwise $T_0'$ would be a clique path
tree of $G$.  Moreover:
%
    \begin{claim}\label{cla}
    For any $a\in A$, the two leaves of $T_0^a$ are in $\mathcal L$ and
    at least one of them is in~$\mathcal L^*$.
    \end{claim}
\begin{proofcl}
Let $L_1, L_2$ be the leaves of $T_0^a$, and, for $i=1, 2$, let
$\ell_i \in L_i\setminus S_{L_i}$.  We have $a\in S_{L_1}$, and $a$ is
not in any member of $\mathcal S(L_1)\setminus S_{L_1}$.  Thus $L_1\in
\mathcal L$.  Similarly $L_2\in \mathcal L$.  The three vertices $q,
\ell_1, \ell_2$ are adjacent to $a$, so they do not form an asteroidal
triple by Lemma~\ref{lem:PAT}, and so one of them is the middle of the
other two.  Vertex $q$ cannot be the middle of $\ell_1, \ell_2$ by
Claim~\ref{clq}.  So we may assume up to symmetry that $\ell_1$ is the
middle of $q, \ell_2$.  So, by Lemma~\ref{lem:middle}, there is an
edge of $T_0'[Q, L_2]$ with a label included in $S_{L_1}$.  So
$T_{L_1}$ is a strict subtree of $T_0'\setminus L_1$ and so
$L_1\in\mathcal L^*$.
\end{proofcl}

The preceding claim implies that $\mathcal L^*$ is not empty.  We
choose $L\in \mathcal L^*$ such that the subtree $T_L$ is maximal.
Let $S_{Q'}$ be the label of the edge of $T_0[L, Q']$ that is incident
to $Q'$.  Vertex $q$ is special and co-special, so there exists $s_Q$
in $S_Q\setminus S_{Q'}$, and we have $s_Q \notin S_L$.  Therefore no
clique of $\mathcal Q\setminus \mathcal Q(T_L)$ contains $s_Q$.  We
add the edge $LL'$ to $T_L$ to obtain a clique tree $T_L'$ of
$G(\mathcal Q(T_L)\cup \{L\})$.  Since $T_L'$ is a strict subtree of
$T_0'$, we can consider a clique path tree $T$ of $G(\mathcal
Q(T_L'))$.  Note that $L$ is a leaf of $T$, for otherwise there are at
least two labels of $T$ that are included in $S_L$, which contradicts
the definition of $T_L$.

\begin{claim}
\label{cl:orange}
Let $a\in A$ be such that both leaves of $T_0^a$ are not in $T_L$.
Let $L_a$ be a leaf of $T_0^a$ that belongs to $\mathcal L^*$.  Then
$L'_a$ is in $T_L$, and every edge $KK'$ of $T_0$ with $K\notin T_L,
K'\in T_L$ satisfies $S_K\subseteq S_{L_a}$.
\end{claim}
\begin{proofcl}
By Claim~\ref{cla}, $L_a$ exists.  Since the labels of the edges of
$T_L$ are not included in $S_L$, they are also not included in
$S_{L_a}$.  So $T_L$ is a subtree of $T_{L_a}$.  By the maximality of
$T_L$, we have $T_L=T_{L_a}$.  By Claim~\ref{claim:L'}, $L'_a$ is in
$T_L$.  By the definition of $T_{L_a}$, every edge $KK'$ of $T_0$ with
$K\notin T_L, K'\in T_L$ satisfies $S_K\subseteq S_{L_a}$.
\end{proofcl}

\begin{claim}
\label{claim:p}
There exist $U, W\in \mathcal Q\setminus \mathcal Q(T_L')$ such that
$UL$ is an edge of $T_0$, $S_U\setminus Q'\neq \emptyset$, $U\cap
W\neq\emptyset$, $W' \in \mathcal Q(T_L)$ and $W\cap Q \neq\emptyset$.
\end{claim}
\begin{proofcl}
We define sets $\mathcal U, \mathcal V$ as follows:
\begin{eqnarray*}
\mathcal U &=& \{U\in\mathcal Q\setminus \mathcal Q(T_L') \mid UL
\textrm{ is an edge of } T_0\}\\
\mathcal V &=& \{V\in\mathcal Q\setminus \mathcal Q(T_L') \mid V' \in
\mathcal Q(T_L)\}.
\end{eqnarray*}

We observe that the members of $\mathcal V$ are pairwise disjoint.
For if there is a vertex $v$ in $V_1\cap V_2$ for some $V_1, V_2
\in\mathcal V$, then $v$ is on three labels (namely $S_{V_1}, S_{V_2}$
and $S_L$) of $T_0$ that do not lie on a common path, contradicting
that $T_0$ is a clique path tree.

We define sets $\mathcal U_p$ ($p \geq 1$) and $\mathcal V_p$ ($p\geq
0$) as follows:
\begin{eqnarray*}
\mathcal V_0 &=& \{W\in\mathcal V \mid W\cap Q \neq\emptyset\}\\
\mathcal U_p &=& \{U\in\mathcal U\setminus (\mathcal
U_1\cup\cdots\cup\mathcal U_{p-1}) \mid \exists\ V\in \mathcal
V_{p-1} \textrm{ such that } U\cap V \neq \emptyset \}\ (p \geq 1)\\
\mathcal V_p &=& \{V\in\mathcal V\setminus (\mathcal
V_1\cup\cdots\cup\mathcal V_{p-1})\ |\ \exists\ U\in \mathcal U_p\
\textrm{such that}\ V\cap U \neq \emptyset \}\ (p \geq 1).
\end{eqnarray*}

Consider the smallest $k\geq 1$ such that there exists $U\in \mathcal
U_k$ with $S_U\setminus Q'\neq \emptyset$.  If no such $U$ exists,
then let $k=\infty$.  The claim states that $k=1$, so let us suppose
on the contrary that $k\ge 2$.  For all $1\leq p\leq k-1$ and all
$U\in \mathcal U_p$, we have $S_U \subseteq Q'$; let $U''\in\mathcal
Q(T)$ be such that $U''\cap S_U\neq \emptyset$ and the length of $T[L,
U'']$ is maximal.  Remark that $S_U$ is included in $U''$ if and only
if all members of $\mathcal Q(T)$ that intersect $S_U$ contain $S_U$.
Let us prove that:
\begin{equation}\label{u2nde}
\mbox{$S_U\subseteq U''$ for every $U\in \mathcal U_p$, $1\leq p\leq
k-1$.}
\end{equation}

Suppose that there exists $U_p\in \mathcal U_p$, $1\leq p\leq k-1$,
such that $S_{U_p}\nsubseteq U_p''$, and let $p$ be minimum with this
property.  Let $V_0, \ldots, V_{p-1}, U_1, \ldots, U_p$ be such that
$V_i\in\mathcal V_i$, $U_i\in\mathcal U_i$, $V_{i-1}\cap U_i \neq
\emptyset$ and $U_i\cap V_i\neq \emptyset$.  Pick $u_i\in U_i
\setminus S_{U_i}$ and $v_i \in V_i \setminus S_{V_i}$.  Let $x_1,
\ldots, x_r$ be such that $x_1\in V_0\cap U_1$, $x_2\in U_1\cap V_1$,
\ldots, $x_r \in V_{p-1}\cap U_p$ with $r=2p-1$.  We claim that $V'_0=
V'_1=\cdots=V'_{p-1}$.  For otherwise there exists $i\in\{1, \ldots,
p-1\}$ such that $V'_{i-1}\neq V'_i$.  Then one of $V'_{i-1}, V'_i$
contains elements of $S_{U_i}$ but not all, and so $S_{U_i}\nsubseteq
U''_i$, which contradicts the minimality of $p$.

By the definition of the $\mathcal V_i$'s, none of $x_2, \ldots, x_r$
is in $Q$.  Let $x_0\in V_0\cap Q$ (maybe $x_0=x_1$).  So $x_0\in
S_{V_0} \subseteq S_L\subset L$.  None of $U_2, \ldots, U_p$ can
contain $x_0$ by the definition of $\mathcal U_1$.  Note that $x_r$ is
in $U_p$ and $V'_{p-1}=V'_0$; on the other hand we have $S_{U_p}
\nsubseteq U_p''$.  So there exists a clique $Z$ of $T_L$ such that
$Z'\in T_0^{x_0}$, $S_{U_p}\subseteq Z'$, $S_{U_p}\cap Z\neq
\emptyset$ and $S_{U_p} \setminus Z\neq \emptyset$.  Vertex $Q'$ is on
$T[L, Z']$ as $S_{U_p}\subseteq Q'$.  Let $z\in Z \setminus Z'$.  We
can find vertices $y_1, \ldots, y_t$ on the labels of $T_0'[Z, Q]$
such that none of them is in $S_L$ and $z$-$y_1$-$\cdots$-$y_t$-$q$ is
a chordless path in $G$.  Let $\ell\in L\setminus S_L$.  By
Claim~\ref{clq}, there exists a path $P$ between $z$ and $\ell$ whose
vertices are not neighbors of $q$.

If $Z\in T_0^{x_0}$, then let $b\in S_{U_p}\setminus Z$.  As $q$ is
special and co-special, we have $S_Q\nsubseteq S_Z$, so let $c\in
S_Q\setminus S_Z$.  Then $z, \ell, q$ form an asteroidal triple
(because of paths $P$, $z$-$y_1$-$\cdots$-$y_t$-$q$ and
$\ell$-$b$-$c$-$q$), and they lie in the neighborhood of $x_0$, a
contradiction.  So $Z\notin T_0^{x_0}$.  Let $x_{r+1}\in Z\cap U_p$.
If $x_{r+1}\in Q$, then $z, \ell, q$ form an asteroidal triple
(because of paths $P$, $z$-$y_1$-$\cdots$-$y_t$-$q$ and
$\ell$-$x_0$-$q$), and they lie in the neighborhood of $x_{r+1}$, a
contradiction.  So $x_{r+1}\notin Q$.  The $S_{U_i}$'s are all
included in $Q'$ and so in $S_L$ too.  They are pairwise disjoint, for
otherwise $T_0$ is not a clique path tree.  Vertex $\ell$ is not in
any of the $S_{U_i}$'s, and $\ell$ is adjacent to all of $x_0, \ldots,
x_{r+1}$ and to none of $u_1, \ldots, u_p, v_0, \ldots, v_{p-1}, y_1,
\ldots, y_t, z, q$.

Suppose that $V_0\cap U_1\cap Q\neq\emptyset$.  Then we may assume
that $x_0=x_1$, so $x_0$ is in $A$ and the two leaves of $T_0^{x_0}$
are not in $T_L$.  By Claim~\ref{cl:orange}, the leaf $L_{x_0}$ of
$T_0^{x_0}$ that is in $\mathcal L^*$ is such that $L'_{x_0}$ is in
$T_L$, so $L_{x_0}=V_0$.  But $x_{r+1}$ is in $Z\cap U_p$, so it is
not in $S_{V_0}$; thus $S_L\nsubseteq S_{V_0}$, which contradicts the
end of Claim~\ref{cl:orange}.  Therefore $V_0\cap U_1\cap Q=
\emptyset$, so $x_0\neq x_1$, $x_0\notin U_1$, $x_1\notin Q$.  Now, if
$t=1$, then $u_1, \ldots, u_p, v_0, \ldots, v_{p-1}, x_0, \ldots,
x_{r+1}, y_1, q, z, \ell$ induce $F_{14}(4p+5)_{p\geq 1}$.  If $t=2$,
then $u_1, \ldots, u_p, v_0, \ldots, v_{p-1}, x_0, \ldots, x_{r+1},
y_1, y_2, q, z, \ell$ induce $F_{15}(4p+6)_{p\geq 1}$.  If $t\geq 3$,
then $\ell, x_0, x_{r+1}, z, y_1, \ldots, y_t, q$ induce
$F_{10}(s+5)_{t\geq 3}$, a contradiction.  Therefore (\ref{u2nde})
holds.

\medskip

Suppose that $k$ is finite.  Let $V_0, \ldots, V_{k-1}, U_1, \ldots,
U_k$ be such that $V_i\in\mathcal V_i$, $U_i\in\mathcal U_i$,
$V_{i-1}\cap U_i \neq \emptyset$, and $U_i\cap V_i\neq \emptyset$.
Let $u_i\in U_i \setminus S_{U_i}$ and $v_i \in V_i \setminus
S_{V_i}$.  Pick vertices $x_1\in V_0\cap U_1$, $x_2\in U_1\cap V_1$,
\ldots, $x_r\in V_{k-1}\cap U_k$ with $r=2k-1$.  By the definition of
$\mathcal V$, none of $x_2, \ldots, x_r$ is in $Q$.  Let $x_0\in
V_0\cap Q$.  Suppose that $V_0\cap U_1\cap Q\neq\emptyset$.  Then we
can assume that $x_0=x_1$, so $x_0$ is in $A$ and the two leaves of
$T_0^{x_0}$ are not in $T_L$.  By Claim~\ref{cl:orange}, the leaf
$L_{x_0}$ of $T_0^{x_0}$ that is in $\mathcal L^*$ is such that
$L'_{x_0}$ is in $T_L$, so $L_{x_0}=V_0$.  But $x_2$ is in $S_{V_1}$
and not in $S_{V_0}$, so $S_{V_1}\nsubseteq S_{V_0}$, which
contradicts the end of Claim~\ref{cl:orange}.  Therefore $V_0\cap
U_1\cap Q=\emptyset$, and $x_0\neq x_1$, $x_0\notin U_1$, $x_1\notin
Q$.  Let $s_{U_k}\in S_{U_k}\setminus Q'$.  Vertex $s_{U_k}$ is not
adjacent to any of $q, s_Q, v_0, \ldots, v_{k-1}$ because
$s_{U_k}\notin Q'$, and by the minimality of $k$, vertex $s_{U_k}$ is
not adjacent to $u_1, \ldots, u_{k-1}$.  Then $u_1, \ldots, u_k, v_0,
\ldots, v_{k-1}, x_0, \ldots, x_{r}, s_{U_k}, s_Q, q$ induce
$F_{16}(4k+3)_{k\geq 2}$, a contradiction.

Now $k$ is infinite.  Then the members of $\bigcup_{p\geq 0} \mathcal
U_p$ are included in $Q'$ and pairwise disjoint, for otherwise $T_0$
is not a clique path tree.  For each member $M$ of $\mathcal U\cup
\mathcal V$, let $T'_0(M)$ be the component of $T_0' \setminus T_L'$
that contains~$M$.  Starting from the path tree $T$ and the trees
$T'_0(M)$ ($M\in \mathcal U\cup \mathcal V$), we build a new tree as
follows.  For each $V\in \bigcup_{p\geq 0} \mathcal V_p$, we add the
edge $VL$ between $T'_0(V)$ and $T$.  For each $U\in \bigcup_{p\geq 1}
\mathcal U_p$, we add the edge $UU''$ between $T'_0(U)$ and $T$.  For
each $U\in \mathcal U\setminus (\bigcup_{p\geq 1} \mathcal U_p)$, we
add the edge $UL$ between $T'_0(U)$ and $T$.  For each $V\in \mathcal
V\setminus (\bigcup_{p\geq 1} \mathcal V_p)$, we define $V''\in
\mathcal Q(T)$ such that $V''\cap S_V\neq\emptyset$ and the length of
$T[L, V'']$ is maximal.  By the definition of $\mathcal V_0$, we have
$S_V\cap Q=\emptyset$, so $V''\neq Q$, so $V''$ is a vertex of $T_L$
on $T_0[{L, V}]$ and it contains $S_V$ as $S_V\subseteq S_L$.  Then we
can add the edge $VV''$ between $T'_0(V)$ and $T$.  Thus we obtain a
clique path tree of $G$, a contradiction.  So $k=1$, and there exist
$U\in \mathcal U_1$ and $W\in \mathcal V_0$ such that $S_U\setminus
Q'\neq\emptyset$, $U\cap W\neq\emptyset$ and $W\cap Q\neq\emptyset$.
\end{proofcl}

Let $U, W$ be as in the preceding claim.  Let $s_U\in S_U\setminus
Q'$.  Vertex $s_U$ is not adjacent to $s_Q$.  Let $u\in U\setminus
S_U$ and $w\in W\setminus S_W$.

\begin{claim}
\label{claim:egal}
    $S_W=S_L$.
\end{claim}
\begin{proofcl}
Assume on the contrary that $S_W\neq S_L$.  Then $S_W$ is a proper
subset of $S_L$.  Suppose that there exists $a\in U\cap W\cap
Q\neq\emptyset$.  Then $a$ is in $A$ and the two leaves of $T_0^{a}$
are not in $T_L$.  By Claim~\ref{cl:orange}, the leaf $L_{a}$ of
$T_0^{a}$ that is in $\mathcal L^*$ is such that $L'_{a}$ is in $T_L$,
so $L_{a}=W$.  But $S_L\nsubseteq S_W$, so Claim~\ref{cl:orange} is
contradicted.  Therefore $U\cap W\cap Q= \emptyset$.  By the
definition of $U$ and $W$, there exists $b\in W\cap Q$ and $c\in U\cap
W$.  So $b\notin U$, $c\notin Q$, $b\neq c$.  Since $s_U$ is in $S_U
\setminus Q'$, we have $S_U\nsubseteq S_W$.  The labels of the edges
of $T_L$ are not included in $S_L$, so they are also not in $S_W$.
Thus we can choose vertices $x_1, \ldots, x_r$ on the labels of
$T_0'[U, Q]$ such that none of the $x_i$'s is in $S_W$, $x_1\in U$,
$x_r\in Q$, and $u$-$x_1$-$\ldots$-$x_r$-$q$ is a path from $u$ to $q$
that avoids $N(w)$.  If $r=1$, then $x_1$ is different from $s_U$ and
$s_Q$, and $w, b, c, u, s_U, x_1, s_Q, q$ induce $F_8$.  If $r=2$,
then, if $x_1$ is adjacent to $s_Q$, vertices $w, b, c, u, s_U, x_1,
s_Q, q$ induce $F_9$, and if $x_1$ is not adjacent to $s_Q$, vertices
$w, b, c, u, x_1, x_2, s_Q, q$ induce $F_9$.  Finally, if $r\geq 3$,
then $w, b, c, u, x_1, \ldots, x_r, q$ induce $F_{10}(r+5)_{r\geq 3}$.
In all cases we obtain a contradiction.
\end{proofcl}

\begin{claim}
\label{claim:W}
$W\in \mathcal L^*$.
\end{claim}

\begin{proofcl}
If $W\in \mathcal L$, then, by Claim~\ref{claim:egal}, we have
$T_W=T_L$ and $W\in \mathcal L^*$, as desired.  So suppose $W\notin
\mathcal L$.  By the definition of $W$, there is a vertex $a\in W\cap
Q$, and so $a\in L$.  Let $L_1, L_2\in \mathcal L$ be the leaves of
$T_0^a$ such that $L_1, L, Q', W, L_2$ lie in this order on that path.
Let $K$ be the member of $\mathcal L$ that is closest to $W$ on
$T_0[L_2, W]$.  Clearly $W\neq K$.  The edges of $T_L$ are not
included in $S_L$, so they are also not in $S_W$ and not in $S_{K}$.
So $T_{K}$ contains $T_L$.  If $K\in \mathcal L^*$, then $T_{K}=T_L$
by the maximality of $T_L$, so $K'\notin T_K$, which contradicts
Claim~\ref{claim:L'}.  Thus $K\notin \mathcal L^*$.  This means that
$T_{K}= T_0'\setminus K$, and so the labels of $T_0'\setminus K$ are
not included in $S_{K}$, in particular $S_W\nsubseteq S_{K}$.  Let
$XX'$ be the edge of $T_0[K, W]$ such that $X'$ contains $S_W$ and $X$
does not (maybe $X'=W$, $X=K$).  The set $S_X$ contains $a$ but not
all of $S_{X'}$, and the members of $\mathcal S_{X'}\setminus
\{S_{X'}, S_X\}$ do not contain $a$.  So no element of $\mathcal
S_{X'}\setminus S_{X'}$ contains $S_{X'}$, which means that $X'\in
\mathcal L$, a contradiction to the definition of $K$.
\end{proofcl}


By Claim~\ref{claim:W}, we have $W\in \mathcal L^*$.  By
Claim~\ref{claim:egal}, we have $T_W=T_L$, so $T_W$ is also maximal
and what we have proved for $L$ can be done for $W$.  Thus, by
Claim~\ref{claim:p}, there exists $X\notin T_W$ such that $XW$ is an
edge of $T_0$ with $S_X\setminus Q'\neq \emptyset$ and $X\cap S_W \neq
\emptyset$.  Let $x\in X\setminus W$ and $s_X\in S_X\setminus Q'$.
Vertex $s_X$ is not in $S_W$, for otherwise it would also be in $S_L$
and in $Q'$.  Vertex $s_U$ is not in $S_L$, for otherwise it would
also be in $S_W$ and in $Q'$.  Vertex $s_Q$ is not in $S_W$ ($=S_L$).
So $s_Q, s_X, s_U$ are pairwise non adjacent.

Suppose that there exists a vertex $a\in U\cap X\cap Q\neq \emptyset$.
So $a\in A$, but none of the two leaves of $T_0^a$ can satisfy
Claim~\ref{cl:orange}, a contradiction.  Therefore $U\cap X\cap Q=
\emptyset$.

Suppose that $U\cap X\neq \emptyset$, and let $a\in U\cap X$.  So $a$
is not in $Q$.  Let $b\in S_W\cap Q$ ($=S_L\cap Q$).  So $b$ is not in
$U\cap X$.  If $b\notin X\cup U$, then $q, u, x, s_Q, s_U, s_X, a, b$
induce $F_6$, a contradiction.  So $b$ is in one of $U, X$, say $b\in
X \setminus U$ (if $b$ is in $U\setminus X$ the argument is similar).
Since $W$ is in $\mathcal L$, there is a vertex $c\in S_W \setminus
S_X$.  Vertex $c$ is adjacent to $a, b, s_U, s_Q$ and not to $x$.
Then $x, a, b, u, s_U, c, s_Q, q$ induce $F_8$, $F_9$ or $F_{10}(8)$,
a contradiction.  Therefore $U\cap X= \emptyset$.

Let $a\in U\cap W$, so $a\notin X$.  Suppose $a\notin Q$.  If there
exists $b\in X\cap Q$, then $b$ is also in $L$ and $q, u, x, s_Q, s_U,
s_X, a, b$ induce $F_6$, a contradiction.  So $X\cap Q=\emptyset$.
Let $c\in W\cap Q$.  Then $c\in L$ and $c\notin X$.  Let $d \in X\cap
S_W$; so $d\in L$, $d\notin Q$, $d\notin U$.  If $c$ is adjacent to
$u$, then $q, u, x, s_Q, s_U, s_X, c, d$ induce $F_6$, else $q, u, x,
s_Q, s_U, s_X, a, c, d$ induce $F_7$, a contradiction.  So $a\in Q$.
Let $e\in X\cap S_W$; so $e\in L$.  If $e\notin Q$, then $q, u, x,
s_Q, s_U, s_X, a, e$ induce $F_6$, a contradiction.  So $e\in Q$.  Let
$f\in S_W\setminus S_Q$ ($f$ exists because $q$ is special and
co-special).  Since $U\cap X=\emptyset$, $f$ is adjacent to at most
one of $u,x$, and then $q, u, x, s_U, s_X, a, e, f$ induce $F_9$ or
$F_{10}(8)$, a contradiction.  This completes the proof of
Theorem~\ref{th:main}.  \hfill $\Box$

\section{Recognition algorithm}

The proof that we give above yields a new recognition algorithm for
path graphs, which takes any graph $G$ as input and either builds a
clique path tree for $G$ or finds one of $F_0, \ldots, F_{16}$.  We
have not analyzed the exact complexity of such a method but it is easy
to see that it is polynomial in the size of the input graph.  More
efficient algorithms were already given by Gavril~\cite{Gav78},
Sch\"affer~\cite{Sch93} and Chaplick~\cite{Cha08}, whose complexity is
respectively $O(n^4)$, $O(nm)$ and $O(nm)$ for graphs with $n$
vertices and $m$ edges.  Another algorithm was proposed in
\cite{DahBai96} and claimed to run in $O(n+m)$ time, but it has only
appeared as an extended abstract (see comments in~\cite[Section
2.1.4]{Cha08}).

There are classical linear time recognition algorithms for
triangulated graphs~\cite{RTL76}, and, following \cite{BL76}, there
have been several linear time recognition algorithms for interval
graphs, of which the most recent is~\cite{HMPV}.  We hope that the
work presented here will be helpful in the search for a linear time
recognition algorithm for path graphs.



\begin{thebibliography}{99}

\bibitem{Ber60}
C. Berge.  Les probl\`emes de coloration en th\'eorie des graphes.
{\it Publ.  Inst.  Stat.  Univ.  Paris} 9 (1960) 123--160.

\bibitem{BL76}
K.S.~Booth, G.S.~Lueker.  Testing for the consecutive ones property,
interval graphs and graph planarity using PQ-tree algorithm.
{\it J.~Comput.  Syst.  Sci.}  13 (1976) 335--379.

\bibitem{Cha08}
S. Chaplick.  {\it PQR-trees and undirected path graphs.}
M.Sc.~Thesis, Dept.  of Computer Science, University of Toronto, 2008.

\bibitem{DahBai96}
E. Dahlhaus, G. Bailey.  Recognition of path graphs in linear time.
5th Italian Conference on Theoretical Computer Science (Revello, 1995)
World Sci.  Publishing, River Edge, NJ, 1996, 201--210.

\bibitem{Dir61}
G.A.~Dirac.  On rigid circuit graphs.  \emph{Abh.  Math.  Sem.  Univ.
Hamburg} 38 (1961) 71--76.

\bibitem{Gav74}
F.~Gavril.  The intersection graphs of subtrees in trees are exactly
the chordal graphs.  \emph{J. Combin.  Theory B} 16 (1974) 47--56.

\bibitem{Gav78}
F.~Gavril.  A recognition algorithm for the intersection graphs of
paths in trees.  \emph{Discrete Math.} 23 (1978) 211--227.

\bibitem{Gol04}
M. C. Golumbic.  {\it Algorithmic graph theory and perfect graphs.}
Annals Disc.  Math.  57, Elsevier, 2004.

\bibitem{HMPV}
M.~Habib, R.~McConnell, C.~Paul, L.~Viennot.  Lex-BFS and partition
refinement, with applications to transitive orientation, interval
graph recognition and consecutive ones testing.  \emph{Theoretical
Computer Science} 234 (2000) 59--84.

\bibitem{HajSur58}
A. Hajnal and J. Sur\'anyi.  \"Uber die Aufl\"osung von Graphen in
vollst\"andige Teilgraphen.  {\it Ann.  Univ.  Sci.  Budapest
E\"otv\"os, Sect.  Math.} 1 (1958) 113--121.


\bibitem{LB62}
C.~Lekkerkerker, D.~Boland.  Representation of finite graphs by a set
of intervals on the real line.  \emph{Fund.  Math.} 51 (1962) 45--64.

\bibitem{MacMac}
T.A. McKee and F.R. McMorris.  {\it Topics in intersection graph
theory.} SIAM Monographs on Discrete Mathematics and Applications,
Philadelphia, 1999.

\bibitem{MW86}
C.L.~Monma, V.K.~Wei.  Intersection graphs of paths in a tree.
\emph{J. Combin.  Theory B} 41 (1986) 141--181.

\bibitem{Pan99}
B. S. Panda.  The forbidden subgraph characterization of directed
vertex graphs.  {\it Discrete Mathematics} 196 (1999) 239--256.

\bibitem{Ren70}
P.L.~Renz.  Intersection representations of graphs by arcs.
\emph{Pacific J. Math.} 34 (1970) 501--510.

\bibitem{RTL76}
D.J.~Rose, R.E.~Tarjan, G.S.~Lueker.  Algorithmic aspects of vertex
elimination of graphs.  \emph{SIAM J. Comput.} 5 (1976) 266--283.


\bibitem{Sch93}
A.A.~Sch\"affer.  A faster algorithm to recognize undirected path
graphs.  \emph{Discrete Appl.  Math.} 43 (1993) 261--295.

\bibitem{TY84}
R.E.~Tarjan, M.~Yannakakis.  Simple linear time algorithms to test
chordality of graphs, test acyclicity of hypergraphs, and selectively
reduce acyclic hypergraphs.  \emph{SIAM J. Comput.} 13 (1984)
566--579.

\bibitem{TonGutSzw05}
S. Tondato, M. Gutierrez, J. Szwarcfiter.  A forbidden subgraph
characterization of path graphs.  {\it Electronic Notes in Discrete
Mathematics} 19 (2005) 281--287.





\end{thebibliography}
\end{document}